\documentstyle{article}
\def\half{\frac{1}{2}}

\def\ol{\overline}
\def\be{\begin{equation}}
\def\ee{\end{equation}}
\def\bea{\begin{eqnarray}}
\def\eea{\end{eqnarray}}
\def\cL{{\cal L}}
\def\cH{{\cal H}}
\def\cF{{\cal F}}
\def\cP{{\cal P}}

\def\mn{{\mu\nu}}
\def\m{\mu}

\def\d{\partial}
\def\bZ{\bar Z}
\def\cY{\bar Y}
\newcommand{\rf}[1]{{\rm (\ref{#1})}}

\begin{document}

\title{Regular Black Holes and Confinement}
\author{Alexander Burinskii\\
Gravity Research Group, NSI Russian
Academy of Sciences\\
B. Tulskaya 52, 115191 Moscow, Russia \\
\&\\ Sergi R. Hildebrandt
\thanks{Temporary address: C/Los Geranios, 8, P210. Candelaria, 38530. S/C.
de Tenerife. Spain. E-mail: hildebrandt@ieec.fcr.es} \\
Institut d'Estudis Espacials de Catalunya (IEEC/CSIC)\\
Edifici Nexus, Gran Capit\`a 2-4, 08034 Barcelona, Spain}
\date{}
\maketitle
\begin{abstract}
\noindent
Properties of the rotating Kerr-Newman black hole solution allow to relate
it with spinning particles.
Singularity of black hole (BH) can be regularized by a metric deformation.
In this case, as a consequence of the Einstein equations,
a material source appears in the form of relativistically
rotating superconducting disk which replaces the former singular region.
We show a relation of the BH regularization to confinement formation.
By regularization, a phase transition occurs near the core of a charged
black hole solution:  from external electrovacuum to an internal
superconducting state of matter.
We discuss two models of such a kind, which demonstrate the appearance of a
baglike structure and a mechanism of confinement based on dual Dirac's
electrodynamics. First one is an approximate
solution based on a supersymmetric charged domain wall, and
second is an exact solution based on nonlinear electrodynamics.
\end{abstract}

\section{Introduction}
Last years there has been a renewed interest in regular black holes (BHs)
and particlelike models within General Relativity (see e.g. \cite{behm}
for a summary), and more especially on electric/magnetically charged BHs
and particlelike solutions (see e.g., \cite{abg,br,bh}).
Singularity of black hole can be regularized by a metric deformation.
In this case, as a consequence of the Einstein equations,
there appears a material source which replaces the former singular region,
and a phase
transition occurs near the core of regularized black hole solutions:  from
external electrovacuum to another state of matter inside the core.
Such regularized black hole solutions can be considered as classical
gravitating solitons.
Regularization of the rotating Kerr-Newman black hole solution is of a
special interest as a solitonic model of charged spinning particle.
We recall that the Kerr-Newman solution has $g=2$ gyromagnetic ratio,
possesses the stringy structures \cite{str} and is a fundamental solution of
low energy string theory. The source of regularized Kerr-Newman
solution takes the form of
relativistically rotating superconducting disk \cite{bag,behm}.

\section{Regularized Kerr-Newman spacetimes}
The Kerr-Newman (KN) solution in the Kerr-Schild (KS) form is:
\be
\label{gkn}
g_{KN} = \eta_{ij} + 2\, h \, k_i \, k_j
\ee
where, in Kerr angular coordinates,
$ \eta_{ij} = {\rm diag}(-1,1,r^2,r^2\,\sin^2\theta) $, $ h $
is a function given by
$ f_{KN}(r)/(r^2+a^2\,\cos^2\theta) $
with $ f_{KN} (r) = m\, r -e^2/2 $ ($ a $, $m $ and $ e $ are the angular
momentum per unit mass, the mass and the electric/magnetic charge of the
source, respectively) and $ k $ is tangent to a very special twisting Kerr
Principal Null Congruence of KN metric that need not be explicitly presented
here.  Looking at~\rf{gkn} a  generalization appears that keeps the
essential
geometry of KN solution: to allow a general $ f(r) $.  In \cite{behm} this
generalization has been studied in detail.

In order to regularize the KN singularity one has to change the behavior of
function $f(r)$ in the region of small $r$ setting $f(r)\sim \alpha r^n $
with $n\le 4$. Particularly, in the nonrotating case, for $n=4$ the source
represents a spherically symmetrical space-time of constant curvature
(de Sitter or anti de Sitter solution) with cosmological
constant $\Lambda= 6 \alpha $ and energy density
$\rho=3 \alpha/ 4\pi$.  The `radial' position $r_0$ of the phase
transition shell can be estimated as a point of intersection for plots
$f_{int}(r)$ and $f_{KN}(r)$, \begin{equation} \frac 4 3 \pi \rho r_0
^4=mr_0-e^2/2.  \label{r0} \end{equation}
Analysis shows \cite{behm} that
there appears a thin intermediate shell at $r=r_0$ with a strong tangential
stress which is typical for domain wall structure.
Dividing Eq. (\ref{r0}) by $r_0$ one can recognize here the
{\it mass balance equation}
$ m=M_{int}(r_0) + M_{em}(r_0)$, where  $M_{int}(r_0)$ is the ADM mass of
the core and
$M_{em}(r_0) =e^2/2r_0$ is external electromagnetic field's one.  As a
consequence of this equation, the AdS and dS interiors are admissible,
and AdS interiors correspond to strongly charged particles.

For rotating BHs, the core region represents a strongly oblated
rotating disk of compton radius, and curvature is concentrated in a
stringy region on the border of this disk. It was shown \cite{bag,sp1} that
there appears a baglike structure in the core region and the source
resembles
a domain wall interpolating between external and internal vacua.  The
Kerr-Schild class of metrics allows one to describe regular rotating and
non-rotating BH solutions at least on the level of gravitational equations.

However, there appear some problems on the level of field models describing
corresponding matter source.
The following demands to the matter field models are necessary for
regularized charged particlelike solutions.

i - External vacuum has to be (super)-Kerr-Newman black hole solution
with {\it long range } electromagnetic field and  zero cosmological
constant;

ii -Internal vacuum has to be an (A)dS space with superconducting
properties, expelling the electric field.

Exact solutions of this kind are unknown and
it is expected that they can exist in some generalized supergravity models.

These demands are very restrictive and are not satisfied in the known
  solitonlike, bag, domain wall and bubble models.
In most known models the external
electromagnetic field is short range. An exception is the $U(I) \times
\tilde U(I) $ field model which was used by Witten to describe
the cosmic superconducting strings \cite{wit}.
This field model can be adapted for description of the superconducting
bag geometry\cite{bag,sp1}.

The model contains two sectors A and B , and correspondingly two Higgs
and
two gauge fields.  One of the gauge fields can be set as a long range
external
electromagnetic field and another one is independent and can be chosen as a
``dual'' gauge field (in the sense of the dual Dirac electrodynamics
\cite{dir}) which is confined inside the bag \cite{bag}.

It is based on the Abelian Higgs field model for superconducting source.
One can achieve nonperturbative soliton-like
solutions of this kind taking the external KN BH field and a core
described by three chiral fields of a supersymmetric field model.
It was shown \cite{bag,sp1} that there exists a
BPS-saturated domain wall solution interpolating between supersymmetric
vacua I) and II). External vacuum I) is characterized by
\begin{equation}
I) \qquad Z=0;\quad \phi=0 ;\quad \vert\sigma\vert=\eta ;\quad W=0,
\label{true}
\end{equation}
and internal vacuum II), used for the interior of the bag, is characterized
by
\begin{equation}
II ) \qquad Z=-m/c;\quad \sigma=0; \quad \vert \phi \vert =\eta
\sqrt{\lambda/c};\quad W=\lambda m\eta ^2/c.
\label{false}
\end{equation}

One can check the existence of the phase transition in the planar wall
approximation.
Minimum of the total energy is achieved by
$r_0=(\frac{e^2}{16\pi \epsilon})^{1/3},$
corresponding to the stationary state with total mass
$M_{tot}^* =E_{tot}^* =\frac {3e^2}{4r_0}.$
For the rotating Kerr-Newman case \cite{bag},  $J=ma$, and for $J\sim 1$ we
find out that parameter $a\sim 1/m$ has Compton size.
Coordinate $r$ is an oblate spheroidal coordinate, and the phase transition
occurs at the shell representing an oblate rotating ellipsoid with the axis
ratio of order $ \sim 137^{-1}$.
\section{Regular BHs based on the models of non-linear electrodynamics}

\subsection{Exact nonrotating regular BH solutions}

Second model demonstrating the phase transition in the core region and
confinement is based on nonlinear electrodynamics of a general (non-BI)
type(NED) \cite{bh}.  It is of a special interests since, at least in the
non-rotating case, it yields the exact and consistent electric/magnetic
charged solutions which can be interpreted as non-perturbative particlelike
solutions.  In spite of the very different field mechanism, this models
displays also a ``dual'' electromagnetic phase in the core region.

In a recent paper
\cite{br}, K. Bronnikov, showed that only purely magnetically charged BHs
could be regular if Maxwell limit was to be reproduced for low energies of
the field. It was showed in \cite{bh} that Bronnikov's restrictions
may be circumvented and indeed both regular magnetically and electrically
charged solutions can exist.

In the obtained recently by Ay\'on-Beato and Garc\'\i a exact  regular
BH solution to this NED-theory, Bronnikov has observed existence of
casps and branches  in Lagrangian. In particular, their solution
contained singular distributions of electric charges $J^i_{(e)}= \nabla_m
F^{mi}$ near a sphere of finite radius $r=r_0 $ and at the center.
In our paper \cite{bh} a modification of this solution was suggested,
which allows one to avoid abnormal branches in the
Ay\'on-Beato -- Garc\'\i a  solution. It was achieved by a transition
to "dual" electrodynamics in the core region (by $r<r_0$)
fulfilling Dirac's idea on dual electrodynamics \cite{dir}.
As a result,
the charges inside the sphere $r=r_0$ turn out to be magnetic $J^\mu _{(m)}=
\nabla_\nu \star \tilde F^{\nu \mu}$.
These charges give rise to a magnetic vacuum
polarization inside the core, while for an external observer this solution
has electric charge only.  On the other hand, the fact that central region
of
this solution contains only magnetic charges allows one to avoid the
Bronnikov theorem and to get regular BH solution with the resulting
electrical charge.

Since core expels electric charges from itself one
can speculate that it possesses superconducting properties, whereas the
vacuum
region of the external observer possesses the dual properties ---it is not
penetrable for magnetic charges, due to their confinement inside the sphere
$r=r_0$.  It is remarkable that nonlinear electrodynamics allows one to get
{\it exact and selfconsistent} solutions of this sort belonging to the
Kerr-Schild class of metrics.

This class of solutions deserves interest not only as an explicit
example of the regular electrically charged BH solution, but also as an
example of particlelike solution with a very specific realization of the
ideas on quark confinement based on dual electrodynamics and
superconductivity \cite{dual}.

\subsection{Regular rotating black holes from NED, preliminary treatment}
The models of rotating regular BH  solutions are of big interest to
modelling spinning gravitating solitons.
In this case it is expected that the
phase transition will occur in a disklike region of compton size, and
there appear a stringy region on the boarder of the Kerr disk.
There was an assumption that such solution in NED-theory can contain a
closed loop of gauge field  similarly to the magnetic vortex fluxes of usual
superconductors.

In this case it is necessary to include, besides $ f= F_{\mn}F^{\mn} $
the other invariant $ g= F_{\mn}{}^\star\! F^{\mn} $.
Contrary to the nonrotating case, the solutions of this kind in Kerr
geometry met obstacles and represent a hard and still unsolved problem.
It is known that the NED theory of the Born-Infeld type cannot lead to
regular BH solutions.  In fact, to be able to solve NED field equations, one
needs to suppose some particular family of spacetimes.  Therefore, it seems
easier to choose some preferred candidates and try to derive the associated
Lagrangian.
Working in the Kerr-Schild class, we have made two relevant advances in such
direction.  First, we have computed the geometry of the most general
spacetime that preserves staticity, axial symmetry and the Kerr
Principal Null
Congruence of KN spacetime. This is accomplished by allowing $ f=f_{int}$
in~\rf{gkn} to depend on $ r $ and $ \theta $ for the short range.
The inclusion of $ \theta $ seems to be essential in the rotating case.  On
the other hand, in the framework of theory with two invariants, the Bianchi
identities are \be \nabla_\m \star F^\mn =0, \label{Bi} \ee and the field
equations take the form \be \nabla_\m ( L_{f} F^\mn + L_{g} \star F^\mn )
=0.
\label{Feq} \ee where $L_{f} = \d{\cL}/\d f, \quad L_{g} = \d{\cL}/\d g$.

These dynamic equations can be expressed via
tensor
$P^\mn = \partial {\cL} /\partial F_\mn = L_{f} F^\mn +L_{g} \star F^\mn,$
in the form
\be
\nabla _\m P^\mn =0;
\label{eqP}
\ee
The following (anti)self-dual combinations can be considered
$\cF_\pm ^\mn =F^\mn \pm i\star F^\mn$,
($\star \cF_\pm ^\mn = \pm i \cF^\mn $)
and
$\cP_\pm ^\mn =P^\mn \pm i\star P^\mn$
($\star \cP_\pm ^\mn = \pm i \cP^\mn $),

One can now introduce two invariants:
$P=P_\mn P^\mn;\quad  Q=P_\mn \star P^\mn$;
and their complex combinations
\be
P_\pm=P\pm iQ =(L_{f} \mp iL_{g})^2(f\pm ig)=4L_{\pm}^2 F_\pm;
\label{Ppm}
\ee
where the following notations are used
\be
F_\pm=f\pm ig; \qquad
L_{\pm}=\d {\cL} /\d F_\pm=\half (L_f\mp i L_g).
\label{Fpm}
\ee
In these terms the following relation holds
\be
\cP_\pm^\mn =2L_{\pm} \cF_\pm^\mn,  \label {cPLF}
\ee
and, correspondingly,
\be
F^\mn = \frac 1{4L_{+}} \cP_+^\mn + \frac 1{4L_{-}} \cP_-^\mn.
\label{cPcF}
\ee
Putting
\be
2H_{\pm}=2\d {\cH} / \d P_\pm = (2L_{\pm})^{-1} \ ,
\label{cH}
\ee
one can obtain
the $FP$-duality via the Legendre transformation similarly to the theory
with one invariant.

Electromagnetic invariants can be expressed in the terms of the Kerr
tetrad components
\bea f&=&F_{ab}F^{ab}= -2(F_{12}^2 + F_{34}^2 )= - (\cF_{12}^2 +
\bar \cF_{12}^2),\nonumber\\ g&=&F_{ab}\star F^{ab}= 4iF_{12}F_{34} =
i(\cF_{12}^2 - \bar \cF_{12}^2).
\label{Kfg}
\eea
In the Kerr-Schild tetrad components the field
equations (\ref{eqP}) can be rewritten via selfdual
and anti-selfdual tensors
containing in the Kerr-Schild null tetrad only two (complex) independent
components,
\be
\cP_{12}=\cP_{34}=P_{34}+P_{12}, \quad \cP_{31}=2P_{31},
\quad \cP_{23}=\cP_{14}=\cP_{24}=0,
\label{sdualP}
\ee
In analogue with nonrotating case one expects that the known solution of the
corresponding linear problem $\cF_{ab}$ can be used as a `starting' solution
for tensor $\cP_{ab}$ in nonlinear case.  This solution has the following
explicit form:
\be
\cP_{12}=AZ^2,\quad \cP_{31}=-(AZ)_{,1},
\label{Psol}
\ee
where for the charged rotating BH solution $A=e/p^2$, $p=2^{-1/2}(1+Y\cY)$,
$Z=p/(r+ia \cos \theta)$.
The invariants $P_\pm$ can be expressed in the Kerr tetrad via complex Kerr
radial coordinate $Z$,
\be P_+=P+iQ =-\cP_{12}^2 = - e^2 (Z/p)^4, \quad
P_-=P-iQ= -e^2 (\bZ/p)^4.  \label{PZ} \ee
The system of field equations (\ref{Feq}) takes the form
\be -iI^1= \cP_{12} H_{+,2} + [\ol{ \cP_{12} H_{+,2}
-\cP_{31} H_{+,4}}] =0, \label{1H+} \ee \be -iI^2= -\ol{\cP_{12} H_{+,2}} -
\cP_{12} H_{+,1} +\cP_{31} H_{+,4} =0, \label{2H+} \ee \be -iI^2= \cP_{12}
H_{+,4}-\ol{\cP_{12} H_{+,4}}=0, \label{3H+} \ee \be -iI^4= \cP_{12} H_{+,3}
+\cP_{31} H_{+,2} -[\ol{ \cP_{12} H_{+,3} + \cP_{31} H_{+,2}}] =0.
\label{4H+}
\ee
Substitution of the ``starting'' solution (\ref{Psol})
in (\ref{4H+}) turns it into a system for function $H_+$. Some nontrivial
solutions of this system were obtained.  However, integration showed that
corresponding Hamiltonian $ \cH $ turned out to be complex that cannot be
appropriate.
Thus, the problem of existence for consistent rotating solutions
in the Kerr-Schild class remains still open.

One interesting observation follows from the above analysis.
The field equations (\ref{Feq}) are similar to the corresponding
field equations in axion-dilaton gravity \cite{ax-dil}.
It shows that the factors $L_f$ and $L_g$ play a role which is analogous
to the role of dilaton and axion.  It leads us to the conclusion that
both types of considered models can be joined in a model of
supergravity containing the gauge and Higgs chiral fields as well as the
axion and dilaton fields.

\end{document}